\documentclass[a4paper]{article}

\usepackage{INTERSPEECH2020}

\usepackage{graphicx}
\usepackage{kotex}
\usepackage{multicol}
\usepackage{array}
\usepackage{subcaption}
\usepackage{cite}

\title{Expressive Text-to-Speech using Style Tag}
\name{Minchan Kim$^1$, Sung Jun Cheon$^1$, Byoung Jin Choi$^1$, Jong Jin Kim$^2$, Nam Soo Kim$^1$}
\address{
  $^1$Department of Electrical and Computer Engineering and INMC, \\Seoul National University, Seoul, South Korea \\
  $^2$SK Telecom
  }
\email{\{mckim, sjcheon, bjchoi\}@hi.snu.ac.kr, kimjj.geek@sk.com, nkim@snu.ac.kr}
 
\begin{document}

\maketitle

\begin{abstract}
 As recent text-to-speech~(TTS) systems have been rapidly improved in speech quality and generation speed, many researchers now focus on a more challenging issue: expressive TTS. To control speaking styles, existing expressive TTS models use categorical style index or reference speech as style input. In this work, we propose StyleTagging-TTS~(ST-TTS), a novel expressive TTS model that utilizes a style tag written in natural language. Using a style-tagged TTS dataset and a pre-trained language model, we modeled the relationship between linguistic embedding and speaking style domain, which enables our model to work even with style tags unseen during training. As style tag is written in natural language, it can control speaking style in a more intuitive, interpretable, and scalable way compared with style index or reference speech. In addition, in terms of model architecture, we propose an efficient non-autoregressive~(NAR) TTS architecture with single-stage training. The experimental result shows that ST-TTS outperforms the existing expressive TTS model, Tacotron2-GST in speech quality and expressiveness.
\end{abstract}

\section{Introduction}
 Recently, end-to-end~(e2e) speech synthesis systems have achieved significant improvement and now show almost human-like speech quality. Among e2e speech synthesis systems, autoregressive~(AR) models such as Tacotron2~\cite{shen2018natural} and Transformer TTS~\cite{li2019neural} first showed state-of-the-art performance by exploiting the attention mechanism. However, as AR models suffer from slow generation speed and a lack of stability due to attention failure, non-autoregressive~(NAR) models~\cite{ren2019fastspeech, ren2020fastspeech, kim2020glow, miao2020flow, donahue2020end, vainer2020speedyspeech} have been proposed in recent years. By explicitly modeling the duration of each text unit and generating speech in parallel, NAR models can synthesize speech much faster than real-time with comparable speech quality with AR models. Such development enables e2e speech synthesis systems to be successfully adopted in many real-world applications.
 
 Meanwhile, to imitate the non-phonetic expressions of human utterance, expressive TTS~\cite{lee2017emotional, tits2019exploring, tits2019visualization, wang2018style, skerry2018towards, jia2018transfer} has been studied in line with the development of the e2e TTS systems. In expressive TTS, the speaking style is modeled in a supervised or unsupervised manner and usually combined with an existing TTS architecture.
 In terms of style control, there are two approaches for expressive TTS. Several models exploit the categorical style labels, which indicate speaking styles such as emotions~\cite{lee2017emotional, tits2019exploring, tits2019visualization}. Trained with a style-labeled dataset in a supervised manner, these models can control style using explicit labels. However, there are limitations in the diversity of expression, as these models can only express a few pre-defined styles. On the other hand, some models use reference speech as style input~\cite{wang2018style, skerry2018towards, jia2018transfer}. In these models, style information is extracted from the reference speech using a reference encoder and transferred to the generated speech. Although these methods don't require a labeled dataset and can express unbounded speaking styles, the extracted style information is not intuitive and interpretable. Moreover, choosing a reference speech every time is time-consuming and memory inefficient, which makes these methods less practical.

 In this paper, we introduce a novel style interface for expressive TTS: style tag. Style tag is a short phrase or word representing the style of an utterance, such as emotion, intention, and tone of voice. As style is tagged in the natural language, it is intuitive and interpretable for controlling the speaking style. In addition, we propose Style Tagging TTS~(ST-TTS), which is a novel non-autoregressive expressive TTS model utilizing style tag as a style input. Using a pre-trained language model as an interpreter for transforming natural language to linguistic embedding, ST-TTS can learn the relationship between linguistic embedding and style embedding space. Due to the generalization capabilities of the language model, even a style tag unseen in the training dataset can be expressed by ST-TTS. Besides, ST-TTS also has several advantages in model architecture. To find alignments between text and speech, ST-TTS uses Monotonic Alignment Search~(MAS) algorithm~\cite{kim2020glow} and normalizing flow~(NF)~\cite{dinh2016density, kingma2018glow} as an aligner. As the aligner is jointly optimized with the entire TTS system, ST-TTS can be trained in single-stage, unlike other NAR models such as FastSpeech ~\cite{ren2019fastspeech, ren2020fastspeech} and SpeedySpeech~\cite{vainer2020speedyspeech}. Additionally, at the inference phase, we can optionally use either style tag or reference speech as a style input by building a shared embedding space of reference speech and style tag.
 
 This paper is organized as follows. In Section 2, we first describe the dataset we used. Then we explain the proposed ST-TTS model in section 3, and experiment results are reported in section 4. Lastly, the conclusion and discussion are covered in section 5.

\section{Style Tagging Dataset}
 As existing expressive TTS datasets~\cite{adigwe2018emotional, livingstone2018ryerson, zhou2020seen} have only a few numbers of speaking styles, a dataset with much more various styles is required for our proposed method. We introduce a Korean Style Tagging TTS dataset: FSNR0\footnote{FSNR0 is a part of SMART-TTS corpus, a Korean TTS corpus for an expressive TTS project named SMART-TTS. We are planning to release this corpus as an open-resource.}. FSNR0 consists of \{speech, transcript, style tag\} tuples and the entire transcripts and style tags are written in Korean. The transcripts are recorded by a voice actress expressing given style tags.
 
 In FSNR0, There are 327 style tags including various emotions, intentions, voice tone and speed. Each style tag appears from one to hundreds of times in the whole dataset except for the normal reading book style, which takes up about 25\% of the total data. The entire speech data is about 26 hours long and consists of about 18,700 sentences. Several examples of style tags are shown in Table \ref{table1}. In training, the style tags are augmented by several rules such as switching adverbs and adjectives, and extending keywords to plausible phrase forms. One of the augmented style tag is sampled every iteration.

 \begin{table}[h]
 \caption{Examples of style tag. We additionally provide the translated ones to help understanding.}
 \label{table1}
 \centering
 \begin{tabular}{@{}ll|ll@{}}
 \toprule
 style tag     & translated      & style tag          & translated   \\ \midrule
 다정하게 & with affection  & 매정하게  & heartless  \\
 화가난 듯    & seem angry & 기쁜 듯 & pleased  \\
 씁쓸한 듯 & bitter  &다급하게 & in a hurry   \\
 큰 소리로  & in a loud voice  & 속삭이듯  & whispering   \\
 졸린 듯  & sleepy  & 술취한 듯  & drunken  \\ \bottomrule
 \end{tabular}
 \end{table}

\begin{figure*}
 \centering
 \begin{subfigure}{0.9\columnwidth}
 \includegraphics[width=\columnwidth]{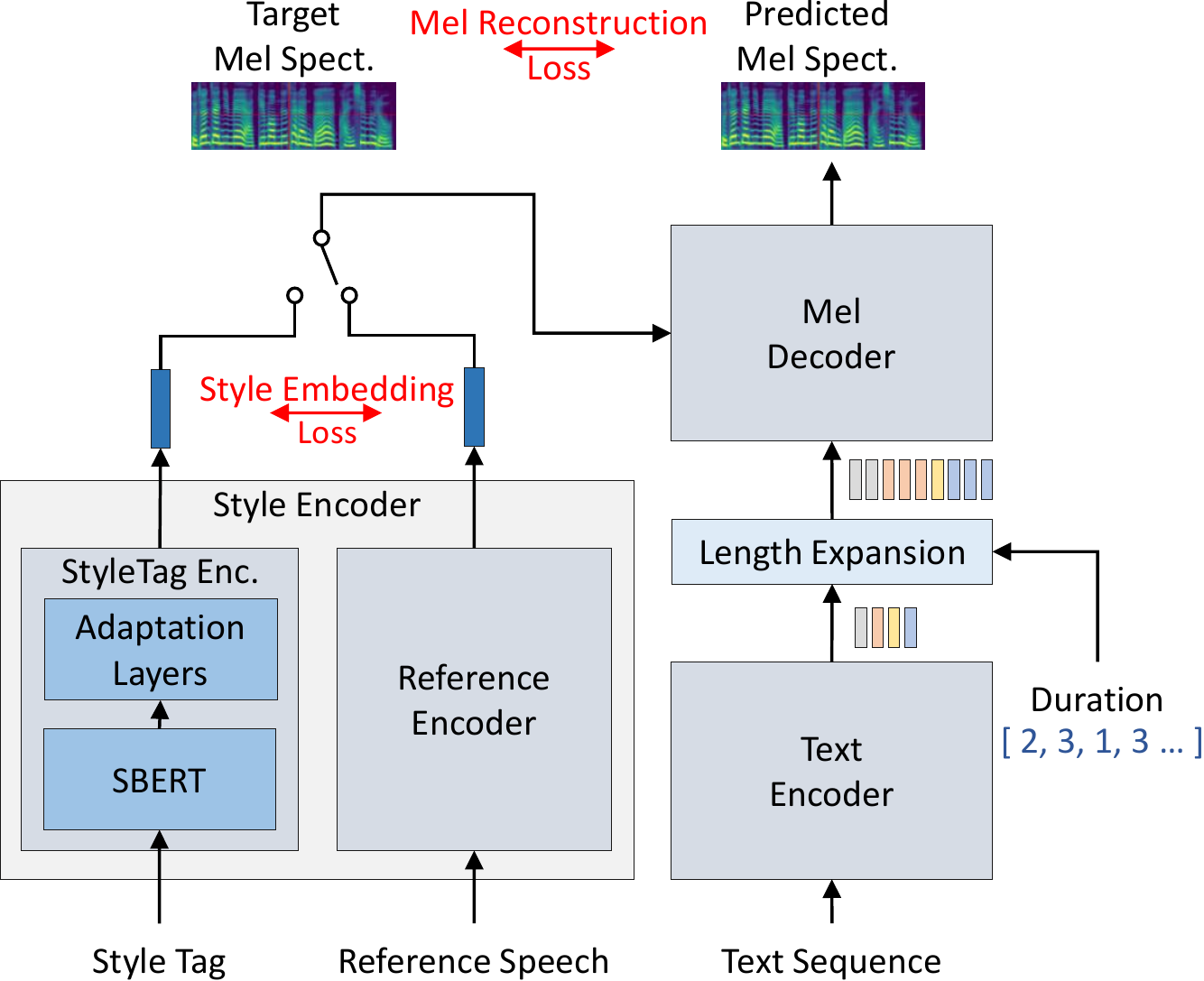}%
 \caption{speech synthesis procedure}%
 \label{subfiga}%
 \end{subfigure}\hspace{1cm}
 \begin{subfigure}{0.9\columnwidth}
 \includegraphics[width=\columnwidth]{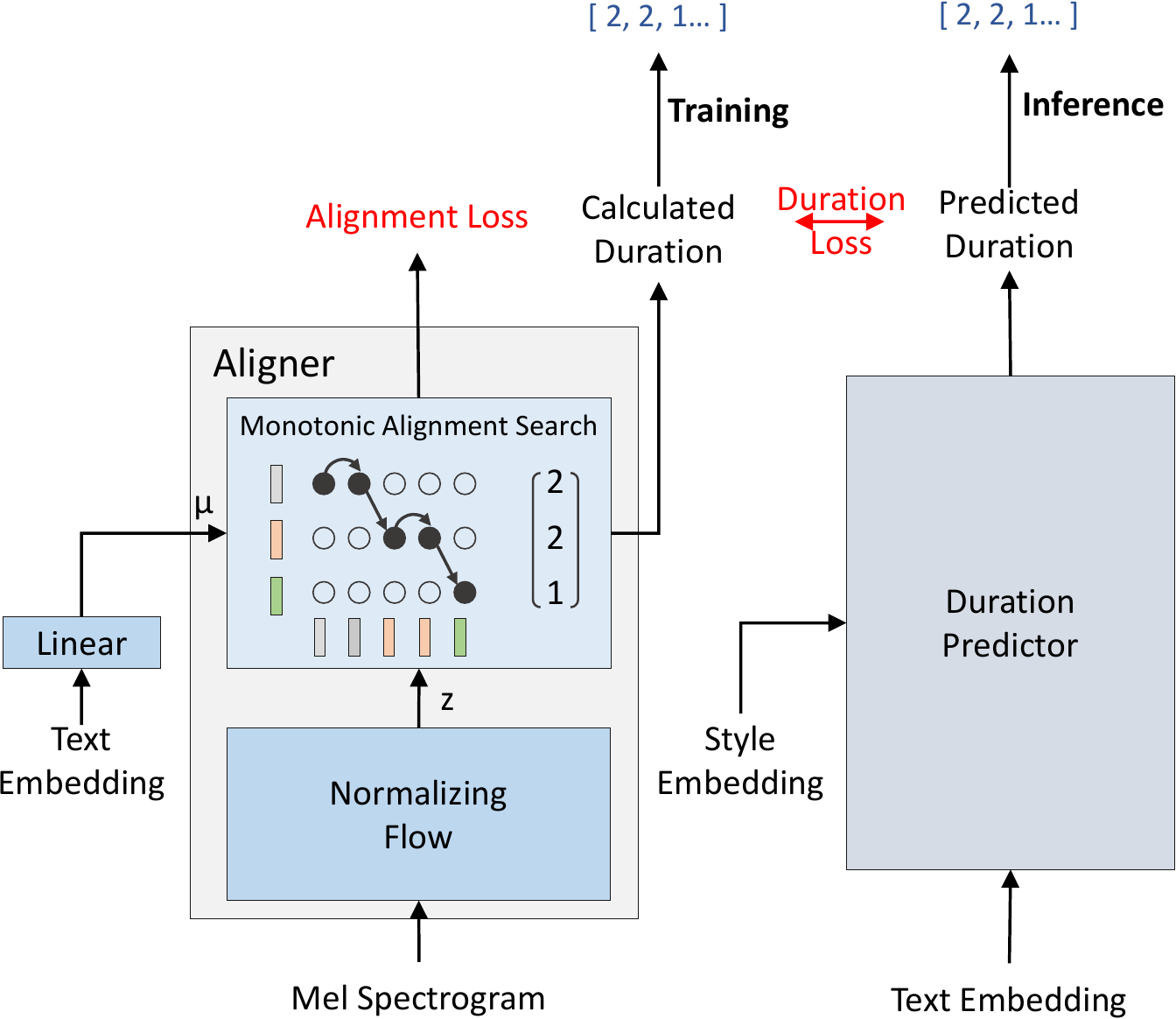}%
 \caption{modules for duration extraction and prediction}%
 \label{subfigb}%
 \end{subfigure}
 \caption{the entire architecture of ST-TTS} 
 \label{figure1}
\end{figure*}

\section{Style Tagging TTS}

\subsection{Model architecture}

 In this section, we introduce the ST-TTS architecture. Our proposed model takes grapheme sequence and style tag as input and returns log mel-spectrogram. Reference speech is additionally used for training and optionally used for inference. The entire structure is based on non-autoregressive feedforward TTS models~\cite{ren2019fastspeech, ren2020fastspeech, vainer2020speedyspeech}, and end-to-end trainable by jointly optimizing duration information with the entire system. The entire architecture is shown in Figure 1, and described as follows.
 
 \subsubsection{Sentence BERT}
 As style tag is an infinite set, we used a pre-trained language model to embed a style tag to a meaningful space without Out-Of-Vocabulary~(OOV) problem. It is the core component in ST-TTS that enables generalization of unseen style tags. We employed Sentence BERT~(SBERT)~\cite{reimers2019sentence} for our proposed model. BERT~\cite{devlin2018bert} is one of the most widely used representation learning for natural language process~(NLP). It is composed of Transformer~\cite{vaswani2017attention} blocks and trained with masking and slot filling task. Although BERT performs well in various NLP tasks such as machine translation, it lacks the ability to extract a sentence-level embedding. SBERT is a fine-tuned model of BERT optimized for sentence-level similarity using contrastive learning. SBERT shows better performance in measuring sentence similarity and clustering sentences than original BERT. We regard style tags as extremely short sentences and used pre-trained SBERT as a component for the style encoder.
 
 \subsubsection{Style encoder}
 Style encoder is a module that extracts style embedding for the TTS system, and it consists of a reference encoder and a style tag encoder. Reference encoder extracts style embedding from reference speech. It has almost the same architecture with reference encoder of Global Style Tokens~(GSTs)~\cite{wang2018style}. The difference is that the proposed module does not use style token, and batch normalization~\cite{ioffe2015batch} layers are replaced with weight normalization~\cite{salimans2016weight}.
 Style tag encoder is a module for extracting a style embedding from a style tag. It consists of pre-trained SBERT and adaptation layers. As mentioned in 3.1.1, SBERT transforms style tag to semantic embedding. Adaptation layers are made up of three linear layers with ReLU activation and learn to map from linguistic semantic space to style embedding space.
 
 In training, style embedding from reference encoder is used for TTS system, and then the embedding from style tag encoder learns reference embedding by mean squared error~(MSE) loss named style embedding loss. This bi-modal conditioning method has the following advantages. 1) By building a bi-modal embedding space of style tag and reference speech, either style tag or reference speech can be used to extract a style embedding for inference. 
 2) Style tag embedding functions as an anchor of reference speeches with the same style tag, which means the reference encoder learns to form clusters of reference speeches with the same style tags. As speeches with similar speaking styles are mapped adjacently in embedding space by the reference encoder, style tags with similar meanings are also located closely in embedding space. This property helps to form a well-defined embedding space, and adaptation layers can easily learn the relationship between style domain and linguistic domain. While training, we freeze SBERT, and the other components are optimized for the entire TTS loss and the style embedding loss.
 
 \subsubsection{Text encoder}
 Text encoder takes a grapheme sequence and returns a text embedding sequence. it consists of residual dilated convolution blocks used in parallel wavegan\footnote{referred code : https://github.com/kan-bayashi/ParallelWaveGAN}~\cite{yamamoto2020parallel}. We used 12 residual blocks with dilation rates~(1, 2, 4) 4 times. The kernel size and hidden dimensions are 5 and 256 respectively.
 
 The output embedding sequence is used for the aligner, duration predictor, and mel decoder. When used for mel decoder, the text embeddings are copied multiple times to match text and speech length using duration information from aligner or duration predictor.

 \subsubsection{Aligner}
 Aligner calculates alignment between text and log mel-spectrogram and returns duration of each grapheme. For aligner, we exploit the method devised in Glow-TTS which uses normalizing flow~(NF) and Monotonic Alignment Search~(MAS) algorithm~\cite{kim2020glow}. To calculate alignment, mel-spectrogram $x_{1:T}$ is first mapped to latent variable $z_{1:T}$ by NF $f^{-1}_{NF}:x\rightarrow z$. Text encoder output is then projected to the latent space with a linear layer and regarded as parameter $\mu_{1:N}$ of the latent variable. $T$ and $N$ denote the length of mel-spectrogram and text tokens respectively. For a possible alignment $A$, $A(j)=i$ represents $z_j$ $\sim\mathcal{N}(\mu_i, \sigma_i$), where $\sigma_i$ of the entire frame is set to 1 for stable training. By change of variables formula used in NF, the conditional log likelihood of mel-spectrogram given text input $c_{1:N}$ and an alignment $A$ can be expressed as Eq.~(1) and Eq.~(2):
 
 \begin{equation}\label{eq:1}
 \log{P_X}(x|c;A) =\log{P_Z}(z|c;A) + \log\left | \det\frac{\partial f_{NF}^{-1}(x)}{\partial x}\right |,
 \end{equation}
 \begin{equation}
 \log{P_Z}(z|c;A)=\sum_{j=1}^{T}\log \mathcal N(z_{j};\mu_{A(j)},\sigma_{A(j)}).
 \end{equation}
 The optimal alignment path $A^*$ for maximizing Eq. (\ref{eq:1}) can be obtained by the MAS algorithm. In training, NF and text encoder are optimized for maximizing $log{P_X}(x|c;A^*)$, and duration for length expansion is calculated by summation $A^*$ over mel-spectrogram axis.
 
 In Glow-TTS, NF is itself a generator of mel-spectrogram, but we only exploit NF as an aligner whose role is just providing an accurate alignment. We used the same architecture with Glow-TTS\footnote{referred code : https://github.com/jaywalnut310/glow-tts} in a much smaller size. The NF is consists of 6 flow blocks, and the affine coupling layers in each flow block are a stack of 4 residual convolution layers with kernel size 5 and hidden dimension 128. Unlike Glow-TTS, we did not squeeze mel-spectrogram for better alignment.
 
 As this aligning method only considers monotonic alignments between text and speech, ST-TTS can be trained much faster and robustly compared with the attention mechanism generally used for TTS alignment. In addition, it does not require auxiliary methods such as guided attention loss~\cite{tachibana2018efficiently} or positional encoding~\cite{vaswani2017attention} for attention mechanism.

 \subsubsection{Duration predictor}
 Duration predictor predicts the number of frames for each grapheme in log scale. Duration predictor is made up of the same residual convolution blocks used in the text encoder. As different style yields different speaking speed, duration predictor also takes a style embedding to predict style dependent duration. Style embedding is duplicated to the equal length of text embedding and conditioned to each residual block by the same method used  in~\cite{yamamoto2020parallel}. Duration predictor is composed of five residual blocks without dilation, and kernel size and hidden dimension are 5 and 256.
 
 \subsubsection{Mel decoder}
 Mel decoder transforms duration-expanded text embedding to log mel-spectrogram conditioned on a style embedding. It has the same architecture with the duration predictor in a much larger size. Mel decoder consists of 30 residual blocks with dilation rate~(1, 2, 4, 8, 16) 6 times. The kernel size used in the mel decoder is 3, and the hidden dimension is 256.

\subsection{Training and inference}

 In training, the total loss is the sum of the following four losses.
 
 \begin{itemize}
 \item Mel reconstruction loss: mean-absolute-error~(MAE) between predicted and target log mel-spectrogram.
 \item Duration loss: huber loss~\cite{huber1992robust} between output of duration predictor and log duration acquired by aligner.
 \item Alignment loss: negative log likelihood loss for training aligner described in 3.1.3
 \item Style embedding loss : MSE loss between style embeddings from reference encoder and style tag encoder as described in 3.1.2
 \end{itemize}
 These losses are summed with equal weights and jointly optimized for the entire model except for SBERT in style encoder.
 
 During inference, ST-TTS uses duration from the duration predictor. As the duration predictor predicts duration in log scale, it is transformed by exponential operation and rounded to the nearest integer. To prevent skipping text units, the predicted durations less than one are clamped to one. We can optionally use style embedding extracted from reference speech or style tag depend on user convenience, as mentioned in 3.1.2.

\section{Experiments}

 \subsection{Experimental setup}
 
 For our experiments, we used the FSNR0 dataset described in section 2. The audio is downsampled from 48 kHz to 22.05 kHz, and 80-dimensional log mel-spectrogram was used for the target and reference speech. For input text, we used Korean grapheme without G2P. Style tags are tokenized by pre-trained SBERT. We trained our model using the Adam optimizier~\cite{kingma2014adam} with learning rate of 0.02 and the Noam learning scheduler~\cite{vaswani2017attention}. Total training took about 87 hours for 500k iteration with a single GeForce GTX 2080 GPU. As a vocoder, we used HiFiGAN~\cite{kong2020hifi} which shows state-of-the-art audio quality for speech synthesis. For the subjective evaluations, 18 native Korean participants joined the tests described in 4.2.1 and 4.2.2. The audio samples of ST-TTS are available on the demo website\footnote{demo site: https://gannnn123.github.io/styletaggingtts-demo}.
 
 \subsection{Evaluation}

 \subsubsection{Audio quality}
 We measure the 5 scale Mean Opinion Score~(MOS) to estimate the speech quality of the proposed model. Since ST-TTS is the first TTS model which can control style with the undefined label or reference speech, we chose Tacotron2-GST~\cite{wang2018style} as a baseline. The result of five groups: ground truth~(GT), GT with vocoder reconstruction, Tacotron2-GST, ST-TTS~(reference speech), ST-TTS~(style tag) is shown in Table 2. ST-TTS~(reference speech) and ST-TTS~(style tag) mean ST-TTS using reference speech and style tag for style input respectively. The reference speech used for Tacotron2-GST and ST-TTS~(reference speech) is the target speech itself because it shows better performance for these models. We tested 30 samples each for normal reading book style samples and expressive samples with style tags, randomly chosen from the test set. According to Table 2, ST-TTS with either style tag or reference speech shows better speech quality than the baseline.

 \begin{table}[h]
 \caption{Mean Opinion Scores with 95\% confidence intervals}
 \centering
 \begin{tabular}{@{}lcc@{}}
 \toprule
 Method            & \multicolumn{1}{l}{Normal} & \multicolumn{1}{l}{Expressive}    \\ \midrule
 GT                & 4.77 $\pm$ 0.04                        & 4.80 $\pm$ 0.04                              \\
 GT~(+ HifiGAN)    & 4.64 $\pm$ 0.05                       & 4.67 $\pm$ 0.05                              \\
 Tacotron2-GST     & 3.37 $\pm$ 0.08                       & 3.44 $\pm$ 0.08                              \\
 ST-TTS~(reference) & 4.15 $\pm$ 0.06                       & 4.12 $\pm$ 0.06                              \\
 ST-TTS~(style tag) & 4.17 $\pm$ 0.06                       & 4.08 $\pm$ 0.06                              \\ \bottomrule
 \end{tabular}
 \end{table}

 \subsubsection{Preference test}
 To demonstrate that the style tag can control the style of generated speech intuitively, we tested a preference test with ST-TTS and Tacotron2-GST. In this test, the participants listened to two samples from ST-TTS and Tacotron2-GST, and rated which sample reflected the given style tag better in 7 scale\footnote{The score is rated in the criterion: A is ( ) than B. Each score is described below.\\-3: much worse, -2: worse, -1: little worse, 0: about same, 1: little better, 2: better, 3: much better}. As Tacotron2-GST cannot directly use the style tag, we used the average style token weights of 10 utterances tagged with the given style tag. We regarded these weights as a representation of the given style tag for Tacotron2-GST. We tested 30 sentences with randomly selected style tags from the test set. The transcripts are sampled from another dataset.
 As a result, ST-TTS got the preference score 1.37~($\pm$0.14, 95\% confidence interval), where the score is ranged in~(-3, 3), and a higher score means better performance. This result shows that ST-TTS can express various speaking styles recognized as reflecting style tags.
 
 \subsubsection{Visualization of style tag embedding}
 In Figure 2, we visualized the style embeddings of style tags using t-SNE~\cite{van2008visualizing}. There are all the style tags in the dataset and several style tags unseen in training. This visualization shows that style tags with similar meanings are embedded adjacently. Particularly, we can see that even the unseen style tags are properly located in the embedding space, which means the available style tag is not limited to the training dataset.

 \begin{figure}[t]
  \centering
\includegraphics[width=\linewidth]{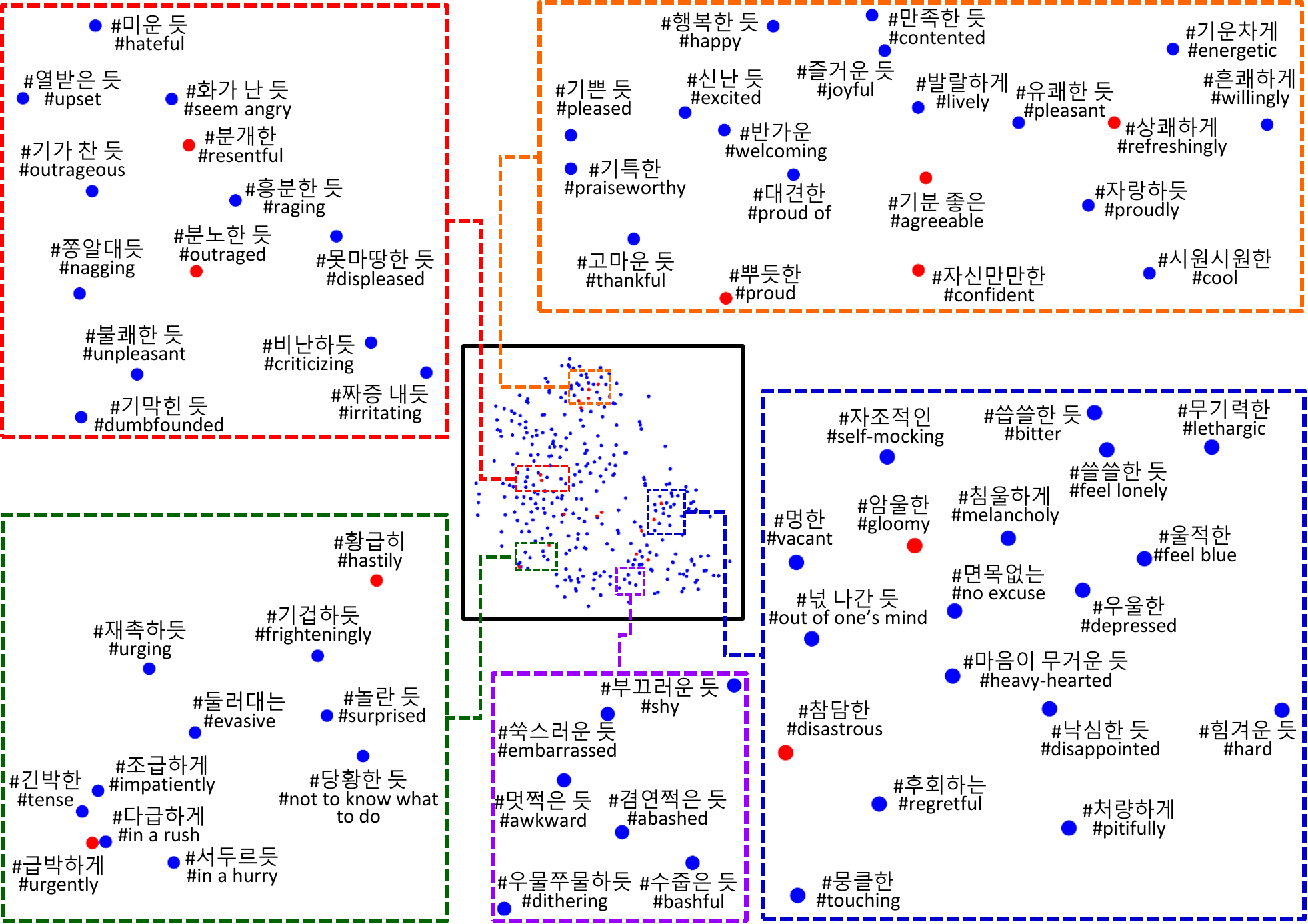}
  \caption{Style embedding visualization using t-SNE. The entire t-SNE plot is at the center, and five local regions are enlarged for closer observation. The blue points and red points indicate the style tags seen and unseen in training, respectively. The translations with similar nuance are provided to help understanding.}
  \label{fig:speech_production}
 \end{figure}

\section{Conclusion and Discussion}
 In this paper, we proposed a novel expressive TTS model named ST-TTS, which uses style tag as a style interface. ST-TTS has several advantages in convenience and performance. 1) As style tag is written in natural language, users can intuitively control the style of generated speech with ST-TTS. The language model, SBERT enables ST-TTS to express even the style tags unseen during training. 2) ST-TTS has a bi-modal embedding space of reference speech and style tag, so either reference style tag or reference speech can be used in inference. 3) ST-TTS is a non-autoregressive TTS model which can be trained in single-stage. For this reason, ST-TTS can be easily trained and has fast generation speed.
  
 As ST-TTS is the first TTS model that uses style tag, there are still remaining research topics as our future works.  
 The most important issue in ST-TTS is the generalization performance for style tags. There are countless style tags in natural language, and the model should express as many style tags as possible for human-level expressiveness. This issue can be addressed in two directions. In terms of data configuration, it is important to use as many style tags as possible in training. We can augment style tags using synonyms to expand the covered linguistic embedding region.
 Next, as style tags are usually much shorter than common sentences, investigating a language model which can embed short phrases better can be a solution for better generalization.
 Meanwhile, there may be styles that cannot be expressed with a single style tag. We are planning to consider the methods using multi-style tags to express various characteristics together. 

\section{Acknowledgements}
 This work was supported by the Institute of Information \& Communications Technology Planning \& Evaluation (IITP) funded by the Korea Government (MSIT) under Grant 2020-0-00059 (Deep learning multi-speaker prosody and emotion cloning technology based on a high quality end-to-end model using small amount of data).

\bibliographystyle{IEEEtran}

\bibliography{mybib}



\end{document}